\documentclass[12pt]{myarticle}

\usepackage{txfonts}
\usepackage[colorlinks]{hyperref}

\newcommand{\n}{\noindent}
\newtheorem{Theorem}{Theorem}

\newtheorem{definition}{Definition}
\newtheorem{ex}{Example}

\begin{document}

\title{F-INDEX OF GRAPHS BASED ON FOUR OPERATIONS RELATED TO THE LEXICOGRAPHIC PRODUCT}

\author{Nilanjan De\corref{cor1}}
\ead{de.nilanjan@rediffmail.com}

\address{Department of
Basic Sciences and Humanities (Mathematics),\\ Calcutta Institute of Engineering and Management, Kolkata, India.}
\cortext[cor1]{Corresponding Author.}

\begin{abstract}
The forgotten topological index or F-index of a graph is defined as the sum of cubes of the degree of all the vertices of the graph. In this paper we study the F-index of four operations related to the lexicographic product on graphs which were introduced by Sarala et al. [D. Sarala, H. Deng, S.K. Ayyaswamya and S. Balachandrana, The Zagreb indices of graphs based on four new operations related to the lexicographic product, \textit{Applied Mathematics and Computation}, \textbf{309} (2017) 156–-169.].

\end{abstract}
\maketitle

\section{Introduction}

Let, $G=(V,E)$ be a connected, undirected simple graph with vertex set $V=V(G)$ and edge set $E=E(G)$. The degree of a vertex $v$ in $G$ is defined as the number of edges incident to $v$ and denoted by ${{d}_{G}}(v)$. In chemical graph theory, chemical structures are considered as a graph, often called molecular graph and a molecular structure descriptor or topological index is a number obtained from a molecular graph and are structurally invariant. Generally, topological indices show a good correlation with different physico-chemical properties of corresponding chemical compounds, so that now a days topological indices are used as a standard tool in studying isomer discrimination and structure-property relations for predicting different properties of chemical compounds and biological activities. Thus, topological indices has shown there applicability in chemistry, biochemistry, nanotechnology and even discovery and design of new drugs. There are various types of topological indices among which the first and second Zagreb indices are most important, most studied and have good correlations to different chemical properties vertex-degree based topological indices. These indices were introduced in 1972 \cite{gutm72}, denoted by $M_1(G)$ and $M_2(G)$ and are respectively defined as

\begin{center}
${{M}_{1}}(G)=\sum\limits_{v\in V(G)}{{{d}_{G}}{{(v)}^{2}}}=\sum\limits_{uv\in E(G)}{[{{d}_{G}}(u)+{{d}_{G}}(v)]}$~~ and~~  ${{M}_{2}}(G)=\sum\limits_{uv\in E(G)}{{{d}_{G}}(u){{d}_{G}}(v)}$.
\end{center}

These indices attracted more and more attention from chemists and mathematicians, specially for different graph operations \cite{kha09,deng16}. Another topological index, named as ``forgotten topological index" or "F-index" \cite{fur15} by Furtula and Gutman is defined as sum of cubes of degrees of the vertices of the graph was also introduced in \cite{gutm72}. Furtula et al., in \cite{fur15a}, investigate some basic properties and bounds of F-index and in \cite{abd15} Abdoa et al. found the extremal trees with respect to the F-index. Recently, the present author studied this index for different graph operations \cite{de16a} and of different classes of nanostar dendrimers \cite{de16c} and also introduced F-coindex in \cite{de16b}. Also, the present author studied F-index of different transformation graphs and four sum of graphs in \cite{de17a} and \cite{de17} respectively. The F-index of a graph $G$ is denoted by $F(G)$, so that

\begin{eqnarray}
F(G)=\sum\limits_{v\in V(G)}{{{d}_{G}}{{(v)}^{3}}}=\sum\limits_{uv\in E(G)}{[{{d}_{G}}{{(u)}^{2}}+{{d}_{G}}{{(v)}^{2}}]}.
\end{eqnarray}

One of the redefined version of Zagreb index denoted by ${Re}Z{{M}}(G)$ and is defined as
\begin{eqnarray}
{Re}Z{{M}}(G)=\sum\limits_{uv\in E(G)}{{{d}_{G}}(u){{d}_{G}}(v)[{{d}_{G}}(u)+{{d}_{G}}(v)]}.
\end{eqnarray}

The general first Zagreb index of a graph $G$ was introduced by Li et al. in \cite{li05} and is defined as
\begin{eqnarray}
{{\xi}_{n}}(G)=\sum\limits_{v\in V(G)}{{{d}_{G}}{{(v)}^{n}}}=\sum\limits_{uv\in E(G)}{[{{d}_{G}}{{(u)}^{n-1}}+{{d}_{G}}{{(v)}^{n-1}}]}
\end{eqnarray}
where $n$ is an integer, not 0 or 1. Obviously ${{\xi }_{2}}(G)=M_1(G)$ and ${{\xi }_{3}}(G)=F(G)$.

There are various subdivision related derived graphs of any graph $G$. For any connected graph $G$, the four derived graphs $S(G)$, $R(G)$, $Q(G)$ and $T(G)$ of $G$ are defined as follows:

(a)  The subdivision graph $S(G)$ is obtained from $G$ by adding a new vertex corresponding to every edge of $G$, that is, each edge of $G$ is replaced by a path of length two.

(b)  The graph $R(G)$ is obtained from $G$ by adding a new vertex corresponding to every edge of $G$,  then joining each new vertex to the end vertices of the corresponding edge that is, each edge of $G$ is replaced by a triangle.

(c)  The graph $Q(G)$ is obtained from $G$ by adding a new vertex corresponding to every edge of $G$, then joining with edges those pairs of new vertices on adjacent edges of $G$.

(a)  The total graph $T(G)$ of a graph $G$  has its vertices as the edges and vertices of $G$ and adjacency in $T(G)$ is defined by the adjacency or incidence of the corresponding elements of $G$.

For different properties and use of the these four derived graphs $S(G)$, $R(G)$, $Q(G)$ and $T(G)$ of $G$, we refer our reader to \cite{nd15,basa16,nd17b,yan07}.

Considering the above four derived graphs, M. Eliasi and B. Taeri introduced four new graph operations named as F-sum graphs in \cite{eli09}, which is based on Cartesian product of graphs.  There are various studies of these F-sum graphs in recent literature \cite{li11,met10,esk13,ana14,deng16,de17}.
Another important type of graph operation, named as the composition or lexicographic product of two connected graphs ${{G}_{1}}$ and ${{G}_{2}}$, denoted by ${{G}_{1}}[{{G}_{2}}]$, is a graph such that the set of vertices is $V({{G}_{1}})\times V({{G}_{2}})$and two vertices $u=({{u}_{1}},{{v}_{1}})$ and  $v=({{u}_{2}},{{v}_{2}})$ of ${{G}_{1}}[{{G}_{2}}]$ are adjacent if and only if either ${{u}_{1}}$is adjacent with ${{u}_{2}}$or ${{u}_{1}}={{u}_{2}}$ and ${{v}_{1}}$ is adjacent with ${{v}_{2}}$.

In \cite{sara17}, Sarala et al. introduced four new operations named as F-product, on these subdivision related graphs based on lexicographic product of two connected graphs ${{G}_{1}}$ and ${{G}_{2}}$ as follows:

\begin{definition} Let $F=\{S,R,Q,T\}$, then the F-product of ${{G}_{1}}$ and ${{G}_{2}}$, denoted by ${{G}_{1}}{{[{{G}_{2}}]}_{F}}$, is defined by $F({{G}_{1}})[{{G}_{2}}]-{{E}^{*}}$, where  ${{E}^{*}}=\{(u,{{v}_{1}})(u,{{v}_{2}})\in E(F({{G}_{1}})[{{G}_{2}}]):u\in V(F({{G}_{1}}))-V({{G}_{1}}),{{v}_{1}}{{v}_{2}}\in E({{G}_{2}})\}$ i.e., ${{G}_{1}}{{[{{G}_{2}}]}_{F}}$ is a graph with the set of vertices $V({{G}_{1}}{{[{{G}_{2}}]}_{F}})=(V({{G}_{1}})\cup E({{G}_{1}}))\times V({{G}_{2}})$ and two vertices $u=({{u}_{1}},{{v}_{1}})$ and $v=({{u}_{2}},{{v}_{2}})$of ${{G}_{1}}[{{G}_{2}}]$ are adjacent if and only if either $[{{u}_{1}}={{u}_{2}}\in V({{G}_{1}})$ and ${{v}_{1}}{{v}_{2}}\in E({{G}_{2}})]$ or $[{{u}_{1}}{{u}_{2}}\in E(F({{G}_{1}}))$ and ${{v}_{1}},{{v}_{2}}\in V({{G}_{2}})]$.
\end{definition}
In \cite{sara17}, Sarala et al. derived explicit expressions of first and second Zagreb indices of F-product graphs.

\section{Main Results}

In this section, if not indicated otherwise, for the graph ${{G}_{i}}$, the notation $V(G_i)$ and ${{E}(G_i)}$ are used for the vertex set and edge set respectively, whereas ${{n}_{i}}$ and ${{m}_{i}}$ denote the number of vertices and the number of edges of the graph ${{G}_{i}}$, $i\in \left\{ 1,2 \right\}$, respectively. In the following we now derive explicit expressions of F-index of the graphs ${{G}_{1}}{{[{{G}_{2}}]}_{S}}$, ${{G}_{1}}{{[{{G}_{2}}]}_{R}}$, ${{G}_{1}}{{[{{G}_{2}}]}_{Q}}$ and ${{G}_{1}}{{[{{G}_{2}}]}_{T}}$ respectively.

\begin{Theorem}
Let ${{G}_{1}}$ and ${{G}_{2}}$ be two connected graphs. Then
\[F({{G}_{1}}{{[{{G}_{2}}]}_{S}})={{n}_{2}}^{4}F({{G}_{1}})+{{n}_{1}}F({{G}_{2}})+6{{n}_{2}}^{2}{{m}_{2}}{{M}_{1}}({{G}_{1}})+6{{n}_{2}}{{m}_{1}}{{M}_{1}}({{G}_{2}})+8{{n}_{2}}^{4}{{m}_{1}}.\]
\end{Theorem}

\n\textit{Proof.} Let, $d(u,v)={{d}_{{{G}_{1}}{{[{{G}_{2}}]}_{S}}}}(u,v)$ be the degree of any vertex $(u,v)$ in the graph ${{G}_{1}}{{[{{G}_{2}}]}_{S}}$. Then from definition of F-index of graph, we have
\begin{eqnarray*}
F({{G}_{1}}{{[{{G}_{2}}]}_{S}})&=&\sum\limits_{({{u}_{1}},{{v}_{1}})({{u}_{2}},{{v}_{2}})\in E({{G}_{1}}{{[{{G}_{2}}]}_{S}})}{[d{{({{u}_{1}},{{v}_{1}})}^{2}}+d{{({{u}_{2}},{{v}_{2}})}^{2}}]}\\
                    &=&\sum\limits_{{{u}_{1}}={{u}_{2}}\in V({{G}_{1}})}{\sum\limits_{{{v}_{1}}{{v}_{2}}\in E({{G}_{2}})}{[d{{({{u}_{1}},{{v}_{1}})}^{2}}+d{{({{u}_{2}},{{v}_{2}})}^{2}}]}}\\
                    &+&\sum\limits_{{{v}_{1}}\in V({{G}_{2}})}{\sum\limits_{{{v}_{2}}\in V({{G}_{2}})}{\sum\limits_{{{u}_{1}}{{u}_{2}}\in E(S({{G}_{1}}))}{[d{{({{u}_{1}},{{v}_{1}})}^{2}}+d{{({{u}_{2}},{{v}_{2}})}^{2}}]}}}\\
                    &=&{{S}_{1}}+{{S}_{2}}
\end{eqnarray*}

Now,
\begin{eqnarray*}
{{S}_{1}}&=&\sum\limits_{{{u}_{1}}={{u}_{2}}\in V({{G}_{1}})}{\sum\limits_{{{v}_{1}}{{v}_{2}}\in E({{G}_{2}})}{[d{{({{u}_{1}},{{v}_{1}})}^{2}}+d{{({{u}_{2}},{{v}_{2}})}^{2}}]}}\\
         &=&\sum\limits_{u\in V({{G}_{1}})}{\sum\limits_{{{v}_{1}}{{v}_{2}}\in E({{G}_{2}})}{[{{\{{{n}_{2}}{{d}_{{{G}_{1}}}}(u)+{{d}_{{{G}_{2}}}}({{v}_{1}})\}}^{2}}+{{\{{{n}_{2}}{{d}_{{{G}_{1}}}}(u)+{{d}_{{{G}_{2}}}}({{v}_{2}})\}}^{2}}]}}\\
         &=&\sum\limits_{u\in V({{G}_{1}})}{\sum\limits_{{{v}_{1}}{{v}_{2}}\in E({{G}_{2}})}{[{{n}_{2}}^{2}{{d}_{{{G}_{1}}}}{{(u)}^{2}}+{{d}_{{{G}_{2}}}}{{({{v}_{1}})}^{2}}+2{{n}_{2}}{{d}_{{{G}_{1}}}}(u){{d}_{{{G}_{2}}}}({{v}_{1}})}}\\
               &+&{{n}_{2}}^{2}{{d}_{{{G}_{1}}}}{{(u)}^{2}}+{{d}_{{{G}_{2}}}}{{({{v}_{2}})}^{2}}+2{{n}_{2}}{{d}_{{{G}_{1}}}}(u){{d}_{{{G}_{2}}}}({{v}_{2}})]\\
         &=&\sum\limits_{u\in V({{G}_{1}})}{\sum\limits_{{{v}_{1}}{{v}_{2}}\in E({{G}_{2}})}{[2{{n}_{2}}^{2}{{d}_{{{G}_{1}}}}{{(u)}^{2}}+\{{{d}_{{{G}_{2}}}}{{({{v}_{1}})}^{2}}+{{d}_{{{G}_{2}}}}{{({{v}_{2}})}^{2}}\}+2{{n}_{2}}{{d}_{{{G}_{1}}}}(u)\{{{d}_{{{G}_{2}}}}({{v}_{1}})+{{d}_{{{G}_{2}}}}({{v}_{2}})\}]}}\\
         &=&2{{n}_{2}}^{2}{{m}_{2}}{{M}_{1}}({{G}_{1}})+{{n}_{1}}F({{G}_{2}})+4{{n}_{2}}{{m}_{1}}{{M}_{1}}({{G}_{2}}).
\end{eqnarray*}

Again,
\begin{eqnarray*}
{{S}_{2}}&=&\sum\limits_{{{v}_{1}}\in V({{G}_{2}})}{\sum\limits_{{{v}_{2}}\in V({{G}_{2}})}{\sum\limits_{{{u}_{1}}{{u}_{2}}\in E(S({{G}_{1}}))}{[d{{({{u}_{1}},{{v}_{1}})}^{2}}+d{{({{u}_{2}},{{v}_{2}})}^{2}}]}}}\\
               &=&\sum\limits_{{{v}_{1}}\in V({{G}_{2}})}{\sum\limits_{{{v}_{2}}\in V({{G}_{2}})}{\sum\limits_{\scriptstyle
 {{u}_{1}}\in V({{G}_{1}}), a\in V({{G}_{1}})\atop \scriptstyle
\ u\ and\ a\ are\ adjacent }{[d{{(u,{{v}_{1}})}^{2}}+d{{(a,{{v}_{2}})}^{2}}]}}}\\
         &=&\sum\limits_{{{v}_{1}}\in V({{G}_{2}})}{\sum\limits_{{{v}_{2}}\in V({{G}_{2}})}{\sum\limits_{\scriptstyle
 {{u}_{1}}\in V({{G}_{1}}), a\in V({{G}_{1}})\atop \scriptstyle \ u\ and\ a\ are\ adjacent
}{[{{\{{{n}_{2}}{{d}_{{{G}_{1}}}}(u)+{{d}_{{{G}_{2}}}}({{v}_{1}})\}}^{2}}+{{\{2{{n}_{2}}\}}^{2}}]}}}\\
 &=&\sum\limits_{{{v}_{1}}\in V({{G}_{2}})}{\sum\limits_{{{v}_{2}}\in V({{G}_{2}})}{\sum\limits_{{{u}}\in V({{G}_{1}})}{{{d}_{{{G}_{1}}}}(u)[{{n}_{2}}^{2}{{d}_{{{G}_{1}}}}{{(u)}^{2}}+{{d}_{{{G}_{2}}}}{{({{v}_{1}})}^{2}}+2{{n}_{2}}{{d}_{{{G}_{1}}}}(u){{d}_{{{G}_{2}}}}({{v}_{1}})+4{{n}_{2}}^{2}]}}}\\
               &=&\sum\limits_{{{v}_{1}}\in V({{G}_{2}})}{\sum\limits_{{{v}_{2}}\in V({{G}_{2}})}{\sum\limits_{{{u}}\in V({{G}_{1}})}{[{{n}_{2}}^{2}{{d}_{{{G}_{1}}}}{{(u)}^{3}}+{{d}_{{{G}_{1}}}}(u){{d}_{{{G}_{2}}}}{{({{v}_{1}})}^{2}}+2{{n}_{2}}{{d}_{{{G}_{1}}}}{{(u)}^{2}}{{d}_{{{G}_{2}}}}({{v}_{1}})+4{{n}_{2}}^{2}{{d}_{{{G}_{1}}}}(u)]}}}\\
               &=&{{n}_{2}}^{4}F({{G}_{1}})+2{{n}_{2}}{{m}_{1}}{{M}_{1}}({{G}_{2}})+4{{n}_{2}}^{2}{{m}_{2}}{{M}_{1}}({{G}_{1}})+8{{n}_{2}}^{4}{{m}_{1}}.
\end{eqnarray*}
Combining, ${{S}_{1}}$ and ${{S}_{2}}$, we get the desired result as theorem 1. \qed

\begin{Theorem}
Let ${{G}_{1}}$ and ${{G}_{2}}$ be two connected graphs. Then
\[F({{G}_{1}}{{[{{G}_{2}}]}_{R}})=8{{n}_{2}}^{4}F({{G}_{1}})+{{n}_{1}}F({{G}_{2}})+24{{n}_{2}}^{2}{{m}_{2}}{{M}_{1}}({{G}_{1}})+12{{n}_{2}}{{m}_{1}}{{M}_{1}}({{G}_{2}})+8{{n}_{2}}^{4}{{m}_{1}}.\]
\end{Theorem}

\n\textit{Proof.} Let, $d(u,v)={{d}_{{{G}_{1}}{{[{{G}_{2}}]}_{R}}}}(u,v)$ be the degree of any vertex $(u,v)$ in the graph ${{G}_{1}}{{[{{G}_{2}}]}_{R}}$. Then similarly, from definition of F-index of graph, we have
\begin{eqnarray*}
F({{G}_{1}}{{[{{G}_{2}}]}_{R}})&=&\sum\limits_{({{u}_{1}},{{v}_{1}})({{u}_{2}},{{v}_{2}})\in E({{G}_{1}}{{[{{G}_{2}}]}_{R}})}{[d{{({{u}_{1}},{{v}_{1}})}^{2}}+d{{({{u}_{2}},{{v}_{2}})}^{2}}]}\\
                  &=&\sum\limits_{{{u}_{1}}={{u}_{2}}\in V({{G}_{1}})}{\sum\limits_{{{v}_{1}}{{v}_{2}}\in E({{G}_{2}})}{[d{{({{u}_{1}},{{v}_{1}})}^{2}}+d{{({{u}_{2}},{{v}_{2}})}^{2}}]}}\\
                 &&+\sum\limits_{{{v}_{1}}\in V({{G}_{2}})}{\sum\limits_{{{v}_{2}}\in V({{G}_{2}})}{\sum\limits_{{{u}_{1}}{{u}_{2}}\in E(R({{G}_{1}}))}{[d{{({{u}_{1}},{{v}_{1}})}^{2}}+d{{({{u}_{2}},{{v}_{2}})}^{2}}]}}}\\
                  &=&{{R}_{1}}+{{R}_{2}}.
\end{eqnarray*}

Now,
\begin{eqnarray}
\nonumber {{R}_{1}}&=&\sum\limits_{{{u}_{1}}={{u}_{2}}\in V({{G}_{1}})}{\sum\limits_{{{v}_{1}}{{v}_{2}}\in E({{G}_{2}})}{[d{{({{u}_{1}},{{v}_{1}})}^{2}}+d{{({{u}_{2}},{{v}_{2}})}^{2}}]}}\\
       \nonumber  &=&\sum\limits_{{{u}_{1}}={{u}_{2}}\in V({{G}_{1}})}{\sum\limits_{{{v}_{1}}{{v}_{2}}\in E({{G}_{2}})}{[{{\{{{n}_{2}}{{d}_{R({{G}_{1}})}}({{u}_{1}})+{{d}_{{{G}_{2}}}}({{v}_{1}})\}}^{2}}+{{\{{{n}_{2}}{{d}_{R({{G}_{1}})}}({{u}_{1}})+{{d}_{{{G}_{2}}}}({{v}_{2}})\}}^{2}}]}}\\
        \nonumber &=&\sum\limits_{u\in V({{G}_{1}})}{\sum\limits_{{{v}_{1}}{{v}_{2}}\in E({{G}_{2}})}{[{{\{2{{n}_{2}}{{d}_{{{G}_{1}}}}(u)+{{d}_{{{G}_{2}}}}({{v}_{1}})\}}^{2}}+{{\{2{{n}_{2}}{{d}_{{{G}_{1}}}}(u)+{{d}_{{{G}_{2}}}}({{v}_{2}})\}}^{2}}]}}\\
       \nonumber  &=&\sum\limits_{u\in V({{G}_{1}})}{\sum\limits_{{{v}_{1}}{{v}_{2}}\in E({{G}_{2}})}{[4{{n}_{2}}^{2}{{d}_{{{G}_{1}}}}{{(u)}^{2}}+{{d}_{{{G}_{2}}}}{{({{v}_{1}})}^{2}}+4{{n}_{2}}{{d}_{{{G}_{1}}}}(u){{d}_{{{G}_{2}}}}({{v}_{1}})}}\\
       \nonumber  &&+4{{n}_{2}}^{2}{{d}_{{{G}_{1}}}}{{(u)}^{2}}+{{d}_{{{G}_{2}}}}{{({{v}_{2}})}^{2}}+4{{n}_{2}}{{d}_{{{G}_{1}}}}(u){{d}_{{{G}_{2}}}}({{v}_{2}})]\\
      \nonumber   &=&\sum\limits_{u\in V({{G}_{1}})}{\sum\limits_{{{v}_{1}}{{v}_{2}}\in E({{G}_{2}})}{[8{{n}_{2}}^{2}{{d}_{{{G}_{1}}}}{{(u)}^{2}}+\{{{d}_{{{G}_{2}}}}{{({{v}_{1}})}^{2}}+{{d}_{{{G}_{2}}}}{{({{v}_{2}})}^{2}}\}+4{{n}_{2}}{{d}_{{{G}_{1}}}}(u)\{{{d}_{{{G}_{2}}}}({{v}_{1}})+{{d}_{{{G}_{2}}}}({{v}_{1}})\}]}}\\
         &=&8{{n}_{2}}^{2}{{m}_{2}}{{M}_{1}}({{G}_{1}})+{{n}_{1}}F({{G}_{2}})+8{{n}_{2}}{{m}_{1}}{{M}_{1}}({{G}_{2}}).
\end{eqnarray}

Now,
\begin{eqnarray*}
{{R}_{2}}&=&\sum\limits_{{{v}_{1}}\in V({{G}_{2}})}{\sum\limits_{{{v}_{2}}\in V({{G}_{2}})}{\sum\limits_{{{u}_{1}}{{u}_{2}}\in E(R({{G}_{1}}))}{[d{{({{u}_{1}},{{v}_{1}})}^{2}}+d{{({{u}_{2}},{{v}_{2}})}^{2}}]}}}\\
         &=&\sum\limits_{{{v}_{1}}\in V({{G}_{2}})}{\sum\limits_{{{v}_{2}}\in V({{G}_{2}})}{\sum\limits_{{{u}_{1}}{{u}_{2}}\in E({{G}_{1}})}{[d{{({{u}_{1}},{{v}_{1}})}^{2}}+d{{({{u}_{2}},{{v}_{2}})}^{2}}]}}}\\
         &&+\sum\limits_{{{v}_{1}}\in V({{G}_{2}})}{\sum\limits_{{{v}_{2}}\in V({{G}_{2}})}{\sum\limits_{\scriptstyle
 {{u}_{1}}{{u}_{2}}\in E(R({{G}_{1}})) \atop \scriptstyle \ {{u}_{1}}\in V({{G}_{1}}),{{u}_{2}}\in V(R({{G}_{1}}))-V({{G}_{1}})
}{[d{{({{u}_{1}},{{v}_{1}})}^{2}}+d{{({{u}_{2}},{{v}_{2}})}^{2}}]}}}\\
         &=&{{R}_{2}}^{\prime }+{{R}_{2}}^{\prime \prime }.
\end{eqnarray*}

Now,
\begin{eqnarray}
\nonumber
{{R}_{2}}^{\prime }&=&\sum\limits_{{{v}_{1}}\in V({{G}_{2}})}{\sum\limits_{{{v}_{2}}\in V({{G}_{2}})}{\sum\limits_{{{u}_{1}}{{u}_{2}}\in E({{G}_{1}})}{[d{{({{u}_{1}},{{v}_{1}})}^{2}}+d{{({{u}_{2}},{{v}_{2}})}^{2}}]}}}\\
                  \nonumber &=&\sum\limits_{{{v}_{1}}\in V({{G}_{2}})}{\sum\limits_{{{v}_{2}}\in V({{G}_{2}})}{\sum\limits_{{{u}_{1}}{{u}_{2}}\in E({{G}_{1}})}{[{{\{{{n}_{2}}{{d}_{R({{G}_{1}})}}({{u}_{1}})+{{d}_{{{G}_{2}}}}({{v}_{1}})\}}^{2}}+{{\{{{n}_{2}}{{d}_{R({{G}_{1}})}}({{u}_{2}})+{{d}_{{{G}_{2}}}}({{v}_{2}})\}}^{2}}]}}}\\
                  \nonumber &=&\sum\limits_{{{v}_{1}}\in V({{G}_{2}})}{\sum\limits_{{{v}_{2}}\in V({{G}_{2}})}{\sum\limits_{{{u}_{1}}{{u}_{2}}\in E({{G}_{1}})}{[{{\{2{{n}_{2}}{{d}_{{{G}_{1}}}}({{u}_{1}})+{{d}_{{{G}_{2}}}}({{v}_{1}})\}}^{2}}+{{\{2{{n}_{2}}{{d}_{{{G}_{1}}}}({{u}_{2}})+{{d}_{{{G}_{2}}}}({{v}_{2}})\}}^{2}}]}}}\\
                  \nonumber &=&\sum\limits_{{{v}_{1}}\in V({{G}_{2}})}{\sum\limits_{{{v}_{2}}\in V({{G}_{2}})}{\sum\limits_{{{u}_{1}}{{u}_{2}}\in E({{G}_{1}})}{[4{{n}_{2}}^{2}{{d}_{{{G}_{1}}}}{{({{u}_{1}})}^{2}}+{{d}_{{{G}_{2}}}}{{({{v}_{1}})}^{2}}+4{{n}_{2}}{{d}_{{{G}_{1}}}}({{u}_{1}}){{d}_{{{G}_{2}}}}({{v}_{1}})}}}\\
                  \nonumber &&+4{{n}_{2}}^{2}{{d}_{{{G}_{1}}}}{{({{u}_{2}})}^{2}}+{{d}_{{{G}_{2}}}}{{({{v}_{2}})}^{2}}+4{{n}_{2}}{{d}_{{{G}_{1}}}}({{u}_{2}}){{d}_{{{G}_{2}}}}({{v}_{2}})]\\
                  \nonumber &=&\sum\limits_{{{v}_{1}}\in V({{G}_{2}})}{\sum\limits_{{{v}_{2}}\in V({{G}_{2}})}{\sum\limits_{{{u}_{1}}{{u}_{2}}\in E({{G}_{1}})}{[4{{n}_{2}}^{2}\{{{d}_{{{G}_{1}}}}{{({{u}_{1}})}^{2}}+{{d}_{{{G}_{1}}}}{{({{u}_{2}})}^{2}}\}+\{{{d}_{{{G}_{2}}}}{{({{v}_{1}})}^{2}}+{{d}_{{{G}_{2}}}}{{({{v}_{2}})}^{2}}\}}}}\\
                   \nonumber &&+4{{n}_{2}}\{{{d}_{{{G}_{1}}}}({{u}_{1}}){{d}_{{{G}_{2}}}}({{v}_{1}})+{{d}_{{{G}_{1}}}}({{u}_{2}}){{d}_{{{G}_{2}}}}({{v}_{2}})\}]\\
                   &=&4{{n}_{2}}^{4}F({{G}_{1}})+2{{n}_{2}}{{m}_{1}}{{M}_{1}}({{G}_{2}})+8{{n}_{2}}^{2}{{m}_{2}}{{M}_{1}}({{G}_{1}}).
\end{eqnarray}

Similarly,

\begin{eqnarray*}
{{R}_{2}}^{\prime \prime }&=&\sum\limits_{{{v}_{1}}\in V({{G}_{2}})}{\sum\limits_{{{v}_{2}}\in V({{G}_{2}})}{\sum\limits_{\scriptstyle
 {{u}_{1}}{{u}_{2}}\in E(R({{G}_{1}})) \atop \scriptstyle \
 {{u}_{1}}\in V({{G}_{1}}),{{u}_{2}}\in V(R({{G}_{1}}))-V({{G}_{1}})
}{[d{{({{u}_{1}},{{v}_{1}})}^{2}}+d{{({{u}_{2}},{{v}_{2}})}^{2}}]}}}\\
       &=&\sum\limits_{{{v}_{1}}\in V({{G}_{2}})}{\sum\limits_{{{v}_{2}}\in V({{G}_{2}})}{\sum\limits_{\scriptstyle
 {{u}_{1}}{{u}_{2}}\in E(R({{G}_{1}})) \atop \scriptstyle \
 {{u}_{1}}\in V({{G}_{1}}),{{u}_{2}}\in V(R({{G}_{1}}))-V({{G}_{1}})
}{[{{\{{{n}_{2}}{{d}_{R({{G}_{1}})}}({{u}_{1}})+{{d}_{{{G}_{2}}}}({{v}_{1}})\}}^{2}}+{{\{{{n}_{2}}{{d}_{R({{G}_{1}})}}({{u}_{2}})\}}^{2}}]}}}\\
       &=&\sum\limits_{{{v}_{1}}\in V({{G}_{2}})}{\sum\limits_{{{v}_{2}}\in V({{G}_{2}})}{\sum\limits_{\scriptstyle
 {{u}_{1}}{{u}_{2}}\in E(R({{G}_{1}})) \atop \scriptstyle \
 {{u}_{1}}\in V({{G}_{1}}),{{u}_{2}}\in V(R({{G}_{1}}))-V({{G}_{1}})
}{[{{\{2{{n}_{2}}{{d}_{{{G}_{1}}}}({{u}_{1}})+{{d}_{{{G}_{2}}}}({{v}_{1}})\}}^{2}}+{{\{2{{n}_{2}}\}}^{2}}]}}}\\
       &=&\sum\limits_{{{v}_{1}}\in V({{G}_{2}})}{\sum\limits_{{{v}_{2}}\in V({{G}_{2}})}{\sum\limits_{\scriptstyle
 {{u}_{1}}{{u}_{2}}\in E(R({{G}_{1}})) \atop \scriptstyle \
 {{u}_{1}}\in V({{G}_{1}}),{{u}_{2}}\in V(R({{G}_{1}}))-V({{G}_{1}})
}{[4{{n}_{2}}^{2}{{d}_{{{G}_{1}}}}{{({{u}_{1}})}^{2}}+{{d}_{{{G}_{2}}}}{{({{v}_{1}})}^{2}}+4{{n}_{2}}{{d}_{{{G}_{1}}}}({{u}_{1}}){{d}_{{{G}_{2}}}}({{v}_{1}})}}}\\
          &&+4{{n}_{2}}^{2}]\\
      &=&4{{n}_{2}}^{4}\sum\limits_{{{u}_{1}}\in V({{G}_{1}})}{{{d}_{{{G}_{1}}}}{{({{u}_{1}})}^{3}}}+{{n}_{2}}{{M}_{1}}({{G}_{2}})\sum\limits_{{{u}_{1}}\in V({{G}_{1}})}{{{d}_{{{G}_{1}}}}({{u}_{1}})}+8{{n}_{2}}^{2}{{m}_{2}}\sum\limits_{{{u}_{1}}\in V({{G}_{1}})}{{{d}_{{{G}_{1}}}}{{({{u}_{1}})}^{2}}}\\
      &&+4{{n}_{2}}^{4}\sum\limits_{{{u}_{1}}\in V({{G}_{1}})}{{{d}_{{{G}_{1}}}}({{u}_{1}})}\\
      &=&4{{n}_{2}}^{4}F({{G}_{1}})+2{{n}_{2}}{{m}_{1}}{{M}_{1}}({{G}_{2}})+8{{n}_{2}}^{2}{{m}_{2}}{{M}_{1}}({{G}_{1}})+8{{n}_{2}}^{4}{{m}_{1}}.
\end{eqnarray*}
Hence combining the above results we get the desired result.    \qed

\begin{Theorem}
Let ${{G}_{1}}$ and ${{G}_{2}}$ be two connected graphs. Then
\begin{eqnarray*}
F({{G}_{1}}{{[{{G}_{2}}]}_{Q}})&=&{{n}_{1}}F({{G}_{2}})-{{n}_{2}}^{4}F({{G}_{1}})+3{{n}_{2}}^{4}{Re}ZM({{G}_{1}})+2{{n}_{2}}^{4}HM({{G}_{1}})+6{{n}_{2}}^{2}{{m}_{2}}{{M}_{1}}({{G}_{1}})\\
                               &&+6{{n}_{2}}{{m}_{1}}{{M}_{1}}({{G}_{2}})+{{n}_{2}}^{4}{{\xi}_{4}}({{G}_{1}})-4{{n}_{2}}^{4}{{M}_{2}}({{G}_{1}}).
\end{eqnarray*}
\end{Theorem}

\n\textit{Proof.} Let, $d(u,v)={{d}_{{{G}_{1}}{{[{{G}_{2}}]}_{Q}}}}(u,v)$ be the degree of any vertex $(u,v)$ in the graph ${{G}_{1}}{{[{{G}_{2}}]}_{Q}}$. So, from definition of F-index of graph, we can write
\begin{eqnarray*}
F({{G}_{1}}{{[{{G}_{2}}]}_{Q}})&=&\sum\limits_{({{u}_{1}},{{v}_{1}})({{u}_{2}},{{v}_{2}})\in E({{G}_{1}}{{[{{G}_{2}}]}_{Q}})}{[d{{({{u}_{1}},{{v}_{1}})}^{2}}+d{{({{u}_{2}},{{v}_{2}})}^{2}}]}\\
                  &=&\sum\limits_{{{u}_{1}}={{u}_{2}}\in V({{G}_{1}})}{\sum\limits_{{{v}_{1}}{{v}_{2}}\in E({{G}_{2}})}{[d{{({{u}_{1}},{{v}_{1}})}^{2}}+d{{({{u}_{2}},{{v}_{2}})}^{2}}]}}\\
                 &&+\sum\limits_{{{v}_{1}}\in V({{G}_{2}})}{\sum\limits_{{{v}_{2}}\in V({{G}_{2}})}{\sum\limits_{{{u}_{1}}{{u}_{2}}\in E(Q({{G}_{1}}))}{[d{{({{u}_{1}},{{v}_{1}})}^{2}}+d{{({{u}_{2}},{{v}_{2}})}^{2}}]}}}\\
                  &=&{{Q}_{1}}+{{Q}_{2}}.
\end{eqnarray*}
Now,
\begin{eqnarray*}
{{Q}_{1}}&=&\sum\limits_{{{u}_{1}}={{u}_{2}}\in V({{G}_{1}})}{\sum\limits_{{{v}_{1}}{{v}_{2}}\in E({{G}_{2}})}{[d{{({{u}_{1}},{{v}_{1}})}^{2}}+d{{({{u}_{2}},{{v}_{2}})}^{2}}]}}\\
         &=&\sum\limits_{u\in V({{G}_{1}})}{\sum\limits_{{{v}_{1}}{{v}_{2}}\in E({{G}_{2}})}{[{{\{{{n}_{2}}{{d}_{{Q({G}_{1}})}}(u)+{{d}_{{{G}_{2}}}}({{v}_{1}})\}}^{2}}+{{\{{{n}_{2}}{{d}_{{Q({G}_{1})}}}(u)+{{d}_{{{G}_{2}}}}({{v}_{2}})\}}^{2}}]}}\\ \nonumber
         &=&\sum\limits_{u\in V({{G}_{1}})}{\sum\limits_{{{v}_{1}}{{v}_{2}}\in E({{G}_{2}})}{[{{n}_{2}}^{2}{{d}_{{{G}_{1}}}}{{(u)}^{2}}+{{d}_{{{G}_{2}}}}{{({{v}_{1}})}^{2}}+2{{n}_{2}}{{d}_{{{G}_{1}}}}(u){{d}_{{{G}_{2}}}}({{v}_{1}})}}\\
         &&+{{n}_{2}}^{2}{{d}_{{{G}_{1}}}}{{(u)}^{2}}+{{d}_{{{G}_{2}}}}{{({{v}_{2}})}^{2}}+2{{n}_{2}}{{d}_{{{G}_{1}}}}(u){{d}_{{{G}_{2}}}}({{v}_{2}})]\\
         &=&\sum\limits_{u\in V({{G}_{1}})}{\sum\limits_{{{v}_{1}}{{v}_{2}}\in E({{G}_{2}})}{[2{{n}_{2}}^{2}{{d}_{{{G}_{1}}}}{{(u)}^{2}}+\{{{d}_{{{G}_{2}}}}{{({{v}_{1}})}^{2}}+{{d}_{{{G}_{2}}}}{{({{v}_{2}})}^{2}}\}+2{{n}_{2}}{{d}_{{{G}_{1}}}}(u)\{{{d}_{{{G}_{2}}}}({{v}_{1}})+{{d}_{{{G}_{2}}}}({{v}_{2}})\}]}}\\
         &=&2{{n}_{2}}^{2}{{m}_{2}}{{M}_{1}}({{G}_{1}})+{{n}_{1}}F({{G}_{2}})+4{{n}_{2}}{{m}_{1}}{{M}_{1}}({{G}_{2}}).
\end{eqnarray*}

Again,
\begin{eqnarray*}
{{Q}_{2}}&=&\sum\limits_{{{v}_{1}}\in V({{G}_{2}})}{\sum\limits_{{{v}_{2}}\in V({{G}_{2}})}{\sum\limits_{{{u}_{1}}{{u}_{2}}\in E(Q({{G}_{1}}))}{[d{{({{u}_{1}},{{v}_{1}})}^{2}}+d{{({{u}_{2}},{{v}_{2}})}^{2}}]}}}\\
         &=&\sum\limits_{{{v}_{1}}\in V({{G}_{2}})}{\sum\limits_{{{v}_{2}}\in V({{G}_{2}})}{\sum\limits_{\scriptstyle
 {{u}_{1}}{{u}_{2}}\in E(Q({{G}_{1}})) \atop \scriptstyle \
 {{u}_{1}}\in V({{G}_{1}}),{{u}_{2}}\in V(Q({{G}_{1}}))-V({{G}_{1}})
}{[d{{({{u}_{1}},{{v}_{1}})}^{2}}+d{{({{u}_{2}},{{v}_{2}})}^{2}}]}}}\\
        &&+\sum\limits_{{{v}_{1}}\in V({{G}_{2}})}{\sum\limits_{{{v}_{2}}\in V({{G}_{2}})}{\sum\limits_{\scriptstyle
 {{u}_{1}}{{u}_{2}}\in E(Q({{G}_{1}})) \atop \scriptstyle \
 {{u}_{1}},{{u}_{2}}\in V(Q({{G}_{1}}))-V({{G}_{1}})
}{[d{{({{u}_{1}},{{v}_{1}})}^{2}}+d{{({{u}_{2}},{{v}_{2}})}^{2}}]}}}\\
        &=&{{Q}_{2}}^{\prime }+{{Q}_{2}}^{\prime \prime }.
\end{eqnarray*}

Now,

\begin{eqnarray*}
{{Q}_{2}}^{\prime }&=&\sum\limits_{{{v}_{1}}\in V({{G}_{2}})}{\sum\limits_{{{v}_{2}}\in V({{G}_{2}})}{\sum\limits_{\scriptstyle
 {{u}_{1}}{{u}_{2}}\in E(Q({{G}_{1}})) \atop \scriptstyle \
 {{u}_{1}}\in V({{G}_{1}}),{{u}_{2}}\in V(Q({{G}_{1}}))-V({{G}_{1}})
}{[d{{({{u}_{1}},{{v}_{1}})}^{2}}+d{{({{u}_{2}},{{v}_{2}})}^{2}}]}}}\\
                  &=&\sum\limits_{{{v}_{1}}\in V({{G}_{2}})}{\sum\limits_{{{v}_{2}}\in V({{G}_{2}})}{\sum\limits_{\scriptstyle
 {{u}_{1}}{{u}_{2}}\in E(Q({{G}_{1}})) \atop \scriptstyle \
 {{u}_{1}}\in V({{G}_{1}}),{{u}_{2}}\in V(Q({{G}_{1}}))-V({{G}_{1}})
}{[{{\{{{n}_{2}}{{d}_{Q({{G}_{1}})}}({{u}_{1}})+{{d}_{{{G}_{2}}}}({{v}_{1}})\}}^{2}}+{{\{{{n}_{2}}{{d}_{Q({{G}_{1}})}}({{u}_{2}})\}}^{2}}]}}}\\
                  &=&\sum\limits_{{{v}_{1}}\in V({{G}_{2}})}{\sum\limits_{{{v}_{2}}\in V({{G}_{2}})}{\sum\limits_{\scriptstyle
 {{u}_{1}}{{u}_{2}}\in E(Q({{G}_{1}})) \atop \scriptstyle \
 {{u}_{1}}\in V({{G}_{1}}),{{u}_{2}}\in V(Q({{G}_{1}}))-V({{G}_{1}})
}{[{{n}_{2}}^{2}{{d}_{{{G}_{1}}}}{{({{u}_{1}})}^{2}}+{{d}_{{{G}_{2}}}}{{({{v}_{1}})}^{2}}+2{{n}_{2}}{{d}_{{{G}_{1}}}}({{u}_{1}}){{d}_{{{G}_{2}}}}({{v}_{1}})}}}\\
       &&+{{n}_{2}}^{2}{{d}_{{{G}_{1}}}}{{({{u}_{2}})}^{2}}]\\
                  &=&{{n}_{2}}^{4}F({{G}_{1}})+2{{n}_{2}}{{m}_{1}}{{M}_{1}}({{G}_{2}})+4{{n}_{2}}^{2}{{m}_{2}}{{M}_{1}}({{G}_{1}})\\
                  &&+{{n}_{2}}^{2}\sum\limits_{{{v}_{1}}\in V({{G}_{2}})}{\sum\limits_{{{v}_{2}}\in V({{G}_{2}})}{\sum\limits_{\scriptstyle
 {{u}_{1}}{{u}_{2}}\in E(Q({{G}_{1}})) \atop \scriptstyle \
 {{u}_{1}}\in V({{G}_{1}}),{{u}_{2}}\in V(Q({{G}_{1}}))-V({{G}_{1}})
}{{{d}_{Q({{G}_{1}})}}{{({{u}_{2}})}^{2}}}}}.
\end{eqnarray*}

Now, since ${{d}_{Q({{G}_{1}})}}({{u}_{2}})={{d}_{{{G}_{1}}}}({{w}_{i}})+{{d}_{{{G}_{1}}}}({{w}_{j}})$, for ${{u}_{2}}\in V(Q({{G}_{1}}))-V({{G}_{1}})$, where ${{u}_{2}}$ is the vertex inserted into the edge ${{w}_{i}}{{w}_{j}}$ of ${{G}_{1}}$, we have
\begin{eqnarray}
\nonumber
\sum\limits_{\scriptstyle
 {{u}_{1}}{{u}_{2}}\in E(Q({{G}_{1}})) \atop \scriptstyle \
 {{u}_{1}}\in V({{G}_{1}}),{{u}_{2}}\in V(Q({{G}_{1}}))-V({{G}_{1}})
}{{{d}_{Q({{G}_{1}})}}{{({{u}_{2}})}^{2}}}&=&2\sum\limits_{{{w}_{i}}{{w}_{j}}\in E({{G}_{1}})}{{{[{{d}_{{{G}_{1}}}}({{w}_{i}})+{{d}_{{{G}_{1}}}}({{w}_{j}})]}^{2}}}\\
&=&2HM({{G}_{1}})
\end{eqnarray}

Thus,  ${{Q}_{2}}^{\prime }={{n}_{2}}^{4}F({{G}_{1}})+2{{n}_{2}}{{m}_{1}}{{M}_{1}}({{G}_{2}})+4{{n}_{2}}^{2}{{m}_{2}}{{M}_{1}}({{G}_{1}})+2{{n}_{2}}^{4}HM({{G}_{1}})$.

Again,
\begin{eqnarray}
\nonumber {{Q}_{2}}^{\prime \prime }&=&\sum\limits_{{{v}_{1}}\in V({{G}_{2}})}{\sum\limits_{{{v}_{2}}\in V({{G}_{2}})}{\sum\limits_{\scriptstyle
 {{u}_{1}}{{u}_{2}}\in E(Q({{G}_{1}})) \atop \scriptstyle \
 {{u}_{1}},{{u}_{2}}\in V(Q({{G}_{1}}))-V({{G}_{1}})
}{[d{{({{u}_{1}},{{v}_{1}})}^{2}}+d{{({{u}_{2}},{{v}_{2}})}^{2}}]}}}\\
   \nonumber   &=&\sum\limits_{{{v}_{1}}\in V({{G}_{2}})}{\sum\limits_{{{v}_{2}}\in V({{G}_{2}})}{\sum\limits_{\scriptstyle
 {{u}_{1}}{{u}_{2}}\in E(Q({{G}_{1}})) \atop \scriptstyle \
 {{u}_{1}},{{u}_{2}}\in V(Q({{G}_{1}}))-V({{G}_{1}})
}{[{{\{{{n}_{2}}{{d}_{Q({{G}_{1}})}}({{u}_{1}})\}}^{2}}+{{\{{{n}_{2}}{{d}_{Q({{G}_{1}})}}({{u}_{2}})\}}^{2}}]}}}\\
   \nonumber   &=&\sum\limits_{{{v}_{1}}\in V({{G}_{2}})}{\sum\limits_{{{v}_{2}}\in V({{G}_{2}})}{\sum\limits_{\scriptstyle
 {{u}_{1}}{{u}_{2}}\in E(Q({{G}_{1}})) \atop \scriptstyle \
 {{u}_{1}},{{u}_{2}}\in V(Q({{G}_{1}}))-V({{G}_{1}})
}{{{n}_{2}}^{2}[{{\{{{d}_{{{G}_{1}}}}({{w}_{i}})+{{d}_{{{G}_{1}}}}({{w}_{j}})\}}^{2}}+{{\{{{d}_{{{G}_{1}}}}({{w}_{j}})+{{d}_{{{G}_{1}}}}({{w}_{k}})\}}^{2}}]}}}\\
   \nonumber   &=&{{n}_{2}}^{4}[2\sum\limits_{{{w}_{j}}\in V({{G}_{1}})}{C_{{{d}_{{{G}_{1}}({{w}_{j}})}}}^{2}}\times {{d}_{{{G}_{1}}}}{{({{w}_{j}})}^{2}}+\sum\limits_{{{w}_{j}}\in V({{G}_{1}})}{({{d}_{{{G}_{1}}}}({{w}_{j}})-1)}\sum\limits_{{{w}_{i}}\in V({{G}_{1}}),{{w}_{i}}{{w}_{j}}\in E({{G}_{1}})}{{{d}_{{{G}_{1}}}}{{({{w}_{i}})}^{2}}}\\
   \nonumber    &&+2\sum\limits_{{{w}_{j}}\in V({{G}_{1}})}{{{d}_{{{G}_{1}}}}({{w}_{j}})({{d}_{{{G}_{1}}}}({{w}_{j}})-1)}\sum\limits_{{{w}_{i}}\in V({{G}_{1}}),{{w}_{i}}{{w}_{j}}\in E({{G}_{1}})}{{{d}_{{{G}_{1}}}}({{w}_{i}})}]\\
  \nonumber    &=&{{n}_{2}}^{4}[\sum\limits_{{{w}_{j}}\in V({{G}_{1}})}{\{{{d}_{{{G}_{1}}}}{{({{w}_{j}})}^{4}}-{{d}_{{{G}_{1}}}}{{({{w}_{j}})}^{3}}\}}+\sum\limits_{{{w}_{j}}\in V({{G}_{1}})}{({{d}_{{{G}_{1}}}}({{w}_{j}})-1)}\sum\limits_{{{w}_{i}}\in V({{G}_{1}}),{{w}_{i}}{{w}_{j}}\in E({{G}_{1}})}{{{d}_{{{G}_{1}}}}{{({{w}_{i}})}^{2}}}\\
    \nonumber   &&+2\sum\limits_{{{w}_{j}}\in V({{G}_{1}})}{{{d}_{{{G}_{1}}}}({{w}_{j}})({{d}_{{{G}_{1}}}}({{w}_{j}})-1)}\sum\limits_{{{w}_{i}}\in V({{G}_{1}}),{{w}_{i}}{{w}_{j}}\in E({{G}_{1}})}{{{d}_{{{G}_{1}}}}({{w}_{i}})}]\\
     &=&{{n}_{2}}^{4}[{{\xi}_{4}}({{G}_{1}})-F({{G}_{1}})-4{{M}_{2}}({{G}_{1}})+3{Re}ZM({{G}_{1}})-F({{G}_{1}})].
\end{eqnarray}

Adding the above contributions we get the desired result as theorem 3. \qed

\begin{Theorem}
Let ${{G}_{1}}$ and ${{G}_{2}}$ be two connected graphs. Then
\begin{eqnarray*}
F({{G}_{1}}{{[{{G}_{2}}]}_{T}})&=&{{n}_{1}}F({{G}_{2}})-{{n}_{2}}^{4}F({{G}_{1}})+3{{n}_{2}}^{4}{Re}ZM({{G}_{1}})+2{{n}_{2}}^{4}HM({{G}_{1}})+6{{n}_{2}}^{2}{{m}_{2}}{{M}_{1}}({{G}_{1}})\\
                               &&+6{{n}_{2}}{{m}_{1}}{{M}_{1}}({{G}_{2}})+{{n}_{2}}^{4}{{\xi}_{4}}({{G}_{1}})-4{{n}_{2}}^{4}{{M}_{2}}({{G}_{1}}).
\end{eqnarray*}
\end{Theorem}

\n\textit{Proof.} We have, from definition of total graph $T(G)$

  ${{d}_{{{G}_{1}}{{[{{G}_{2}}]}_{T}}}}(u,v)={{d}_{{{G}_{1}}{{[{{G}_{2}}]}_{R}}}}(u,v)=2{{n}_{2}}{{d}_{{{G}_{1}}}}(u)+{{d}_{{{G}_{2}}}}(v)$, for $u\in V({{G}_{1}})$ and $v\in V({{G}_{2}})$,

  ${{d}_{{{G}_{1}}{{[{{G}_{2}}]}_{T}}}}(u,v)={{d}_{{{G}_{1}}{{[{{G}_{2}}]}_{Q}}}}(u,v)={{n}_{2}}{{d}_{{Q({G}_{1})}}}(u)$, for $u\in V(T({{G}_{1}}))-V({{G}_{1}})$ and $v\in V({{G}_{2}})$.

Then, from the definition of F-index, we have
\begin{eqnarray*}
F({{G}_{1}}{{[{{G}_{2}}]}_{T}})&=&\sum\limits_{({{u}_{1}},{{v}_{1}})({{u}_{2}},{{v}_{2}})\in E({{G}_{1}}{{[{{G}_{2}}]}_{T}})}{[{{d}_{{{G}_{1}}{{[{{G}_{2}}]}_{T}}}}{{({{u}_{1}},{{v}_{1}})}^{2}}+{{d}_{{{G}_{1}}{{[{{G}_{2}}]}_{T}}}}{{({{u}_{2}},{{v}_{2}})}^{2}}]}\\
                     &=&\sum\limits_{u\in V({{G}_{1}})}{\sum\limits_{{{v}_{1}}{{v}_{2}}\in E({{G}_{2}})}{[{{d}_{{{G}_{1}}{{[{{G}_{2}}]}_{T}}}}{{(u,{{v}_{1}})}^{2}}+{{d}_{{{G}_{1}}{{[{{G}_{2}}]}_{T}}}}{{(u,{{v}_{2}})}^{2}}]}}\\
                    &&+\sum\limits_{{{v}_{1}}\in V({{G}_{2}})}{\sum\limits_{{{v}_{2}}\in V({{G}_{2}})}{\sum\limits_{{{u}_{1}}{{u}_{2}}\in E(T({{G}_{1}}))}{[{{d}_{{{G}_{1}}{{[{{G}_{2}}]}_{T}}}}{{({{u}_{1}},{{v}_{1}})}^{2}}+{{d}_{{{G}_{1}}{{[{{G}_{2}}]}_{T}}}}{{({{u}_{2}},{{v}_{2}})}^{2}}]}}}\\
                   &=&\sum\limits_{u\in V({{G}_{1}})}{\sum\limits_{{{v}_{1}}{{v}_{2}}\in E({{G}_{2}})}{[{{d}_{{{G}_{1}}{{[{{G}_{2}}]}_{R}}}}{{(u,{{v}_{1}})}^{2}}+{{d}_{{{G}_{1}}{{[{{G}_{2}}]}_{R}}}}{{(u,{{v}_{2}})}^{2}}]}}\\
                   &&+\sum\limits_{{{v}_{1}}\in V({{G}_{2}})}{\sum\limits_{{{v}_{2}}\in V({{G}_{2}})}{\sum\limits_{{{u}_{1}}{{u}_{2}}\in E({{G}_{1}})}{[{{d}_{{{G}_{1}}{{[{{G}_{2}}]}_{R}}}}{{({{u}_{1}},{{v}_{1}})}^{2}}+{{d}_{{{G}_{1}}{{[{{G}_{2}}]}_{R}}}}{{({{u}_{2}},{{v}_{2}})}^{2}}]}}}\\
                   &&+\sum\limits_{{{v}_{1}}\in V({{G}_{2}})}{\sum\limits_{{{v}_{2}}\in V({{G}_{2}})}{\sum\limits_{\scriptstyle
 {{u}_{1}}{{u}_{2}}\in E(T({{G}_{1}})) \atop \scriptstyle \
 {{u}_{1}}\in V({{G}_{1}}),{{u}_{2}}\in V(T({{G}_{1}}))-V({{G}_{1}})
}{[{{d}_{{{G}_{1}}{{[{{G}_{2}}]}_{R}}}}{{({{u}_{1}},{{v}_{1}})}^{2}}+{{d}_{{{G}_{1}}{{[{{G}_{2}}]}_{Q}}}}{{({{u}_{2}},{{v}_{2}})}^{2}}]}}}\\
                   &&+\sum\limits_{{{v}_{1}}\in V({{G}_{2}})}{\sum\limits_{{{v}_{2}}\in V({{G}_{2}})}{\sum\limits_{\scriptstyle
 {{u}_{1}}{{u}_{2}}\in E(T({{G}_{1}})) \atop \scriptstyle \
 {{u}_{1}},{{u}_{2}}\in V(T({{G}_{1}}))-V({{G}_{1}})
}{[{{d}_{{{G}_{1}}{{[{{G}_{2}}]}_{Q}}}}{{({{u}_{1}},{{v}_{1}})}^{2}}+{{d}_{{{G}_{1}}{{[{{G}_{2}}]}_{Q}}}}{{({{u}_{2}},{{v}_{2}})}^{2}}]}}}
\end{eqnarray*}

Now, we have from (4), (5), (6) and (7)

\[\sum\limits_{u\in V({{G}_{1}})}{\sum\limits_{{{v}_{1}}{{v}_{2}}\in E({{G}_{2}})}{[{{d}_{{{G}_{1}}{{[{{G}_{2}}]}_{R}}}}{{(u,{{v}_{1}})}^{2}}+{{d}_{{{G}_{1}}{{[{{G}_{2}}]}_{R}}}}{{(u,{{v}_{2}})}^{2}}]}}=8{{n}_{2}}^{2}{{m}_{2}}{{M}_{1}}({{G}_{1}})+{{n}_{1}}F({{G}_{2}})+8{{n}_{2}}{{m}_{1}}{{M}_{1}}({{G}_{2}}),\]
and
\begin{eqnarray*}
\sum\limits_{{{v}_{1}}\in V({{G}_{2}})}{\sum\limits_{{{v}_{2}}\in V({{G}_{2}})}{\sum\limits_{{{u}_{1}}{{u}_{2}}\in E({{G}_{1}})}{[{{d}_{{{G}_{1}}{{[{{G}_{2}}]}_{R}}}}{{({{u}_{1}},{{v}_{1}})}^{2}}+{{d}_{{{G}_{1}}{{[{{G}_{2}}]}_{R}}}}{{({{u}_{2}},{{v}_{2}})}^{2}}]}}}\\
=4{{n}_{2}}^{4}F({{G}_{1}})+2{{n}_{2}}{{m}_{1}}{{M}_{1}}({{G}_{2}})+8{{n}_{2}}^{2}{{m}_{2}}{{M}_{1}}({{G}_{1}})
\end{eqnarray*}
and also
\begin{eqnarray*}
&&\sum\limits_{{{v}_{1}}\in V({{G}_{2}})}{\sum\limits_{{{v}_{2}}\in V({{G}_{2}})}{\sum\limits_{\scriptstyle
 {{u}_{1}}{{u}_{2}}\in E(T({{G}_{1}})) \atop \scriptstyle \
 {{u}_{1}}\in V({{G}_{1}}),{{u}_{2}}\in V(T({{G}_{1}}))-V({{G}_{1}})
}{[{{d}_{{{G}_{1}}{{[{{G}_{2}}]}_{R}}}}{{({{u}_{1}},{{v}_{1}})}^{2}}+{{d}_{{{G}_{1}}{{[{{G}_{2}}]}_{Q}}}}{{({{u}_{2}},{{v}_{2}})}^{2}}]}}}\\
&&=\sum\limits_{{{v}_{1}}\in V({{G}_{2}})}{\sum\limits_{{{v}_{2}}\in V({{G}_{2}})}{\sum\limits_{\scriptstyle
 {{u}_{1}}{{u}_{2}}\in E(T({{G}_{1}})) \atop \scriptstyle \
 {{u}_{1}}\in V({{G}_{1}}),{{u}_{2}}\in V(T({{G}_{1}}))-V({{G}_{1}})
}{[{{\{2{{n}_{2}}{{d}_{{{G}_{1}}}}({{u}_{1}})+{{d}_{{{G}_{2}}}}({{v}_{1}})\}}^{2}}+{{\{{{n}_{2}}{{d}_{Q({{G}_{1}})}}({{u}_{2}})\}}^{2}}]}}}\\
&=&\sum\limits_{{{v}_{1}}\in V({{G}_{2}})}{\sum\limits_{{{v}_{2}}\in V({{G}_{2}})}{\sum\limits_{\scriptstyle
 {{u}_{1}}{{u}_{2}}\in E(T({{G}_{1}})) \atop \scriptstyle \
 {{u}_{1}}\in V({{G}_{1}}),{{u}_{2}}\in V(T({{G}_{1}}))-V({{G}_{1}})
}{[4{{n}_{2}}^{2}{{d}_{{{G}_{1}}}}{{({{u}_{1}})}^{2}}+{{d}_{{{G}_{2}}}}{{({{v}_{1}})}^{2}}+4{{n}_{2}}{{d}_{{{G}_{1}}}}({{u}_{1}}){{d}_{{{G}_{2}}}}({{v}_{1}})}}}\\
&&+{{n}_{2}}^{2}{{d}_{Q({{G}_{1}})}}{{({{u}_{2}})}^{2}}]\\
&&=4{{n}_{2}}^{4}F({{G}_{1}})+2{{n}_{2}}{{m}_{1}}{{M}_{1}}({{G}_{2}})+8{{n}_{2}}^{2}{{m}_{2}}{{M}_{1}}({{G}_{1}})+2{{n}_{2}}^{4}HM({{G}_{1}}).
\end{eqnarray*}
Finally
\begin{eqnarray*}
\sum\limits_{{{v}_{1}}\in V({{G}_{2}})}{\sum\limits_{{{v}_{2}}\in V({{G}_{2}})}{\sum\limits_{\scriptstyle
 {{u}_{1}}{{u}_{2}}\in E(T({{G}_{1}})) \atop \scriptstyle \
 {{u}_{1}},{{u}_{2}}\in V(T({{G}_{1}}))-V({{G}_{1}})
}{[{{d}_{{{G}_{1}}{{[{{G}_{2}}]}_{Q}}}}{{({{u}_{1}},{{v}_{1}})}^{2}}+{{d}_{{{G}_{1}}{{[{{G}_{2}}]}_{Q}}}}{{({{u}_{2}},{{v}_{2}})}^{2}}]}}}\\
  ={{n}_{2}}^{4}[{{\xi}_{4}}({{G}_{1}})-2F({{G}_{1}})-4{{M}_{2}}({{G}_{1}})+3{Re}ZM({{G}_{1}})].
\end{eqnarray*}
Now adding the above contributions, we get the desired result.  \qed
\begin{ex}
Let ${{G}_{1}}={{P}_{n}}$ and ${{G}_{2}}={{P}_{m}}$. Then applying Theorems 1-4, for these graphs with ${{n}_{1}}=n$, ${{n}_{2}}=m$,
\begin{eqnarray*}
(i) F({{P}_{n}}{{[{{P}_{m}}]}_{S}})=16n{{m}^{4}}-22{{m}^{4}}+24n{{m}^{3}}-36{{m}^{3}}+12{{m}^{2}}-28nm+36m-14n,
\end{eqnarray*}
\begin{eqnarray*} (ii) F({{P}_{n}}{{[{{P}_{m}}]}_{R}})&=&72n{{m}^{4}}-120{{m}^{4}}+96n{{m}^{3}}-144{{m}^{3}}-48n{{m}^{2}}+96{{m}^{2}}-64nm\\
                                &&+72m-14n,
\end{eqnarray*}
\begin{eqnarray*}
(iii) F({{P}_{n}}{{[{{P}_{m}}]}_{Q}})=72n{{m}^{4}}-152{{m}^{4}}+24n{{m}^{3}}-36{{m}^{3}}+12{{m}^{2}}-28nm+36m-14n,
\end{eqnarray*}
\begin{eqnarray*}
(iv) F({{P}_{n}}{{[{{P}_{m}}]}_{T}})&=&128n{{m}^{4}}-250{{m}^{4}}+96n{{m}^{3}}-144{{m}^{3}}-48n{{m}^{2}}+96{{m}^{2}}-64nm\\
                                    &&+72m-14n.
\end{eqnarray*}
\end{ex}

\section{Conclusion}

In this paper, we derive explicit expression of the forgotten topological index of four new graph operation related to the lexicographic product of graphs denoted by ${{G}_{1}}{{[{{G}_{2}}]}_{S}}$, ${{G}_{1}}{{[{{G}_{2}}]}_{R}}$, ${{G}_{1}}{{[{{G}_{2}}]}_{Q}}$ and ${{G}_{1}}{{[{{G}_{2}}]}_{T}}$ in terms of some other graph invariants such as first and Second Zagreb indices, hyper-Zagreb index, F-index, redefined Zagreb index of the graphs ${{G}_{1}}$ and ${{G}_{2}}$ .

\end{document}